\documentclass[11pt]{article}
\usepackage[a4paper,margin=2.4cm]{geometry}
\usepackage{amsmath,amssymb,amsthm,graphicx,booktabs,lineno,setspace,xcolor}
\usepackage[colorlinks=true,linkcolor=blue!50!black,citecolor=blue!50!black,urlcolor=blue!50!black]{hyperref}
\usepackage[numbers,sort&compress,super]{natbib}
\usepackage{caption}
\newtheorem{conjecture}{Conjecture}
\newcommand{\tr}{\mathrm{tr}}
\title{Noise threshold for learning with quantum memory}
\author{
	Miaojiang Chen\textsuperscript{1,2}, Wenjing Xiao\textsuperscript{1,2,*}, 
	Haiqiang Chen\textsuperscript{1,2}, Jiarong Liang\textsuperscript{1,2}, \\
	Anfeng Liu\textsuperscript{3}, Zhenrong Zhang\textsuperscript{1,2}, 
	Athanasios V. Vasilakos\textsuperscript{4}, Saif Al-Kuwari\textsuperscript{5}, \\
	Ahmed Farouk\textsuperscript{6}, H. Herbert Song\textsuperscript{7,*}
	\thanks{%
		\noindent\textsuperscript{1}School of Computer, Electronics and Information, Guangxi University, Nanning 530004, China.\\
		\textsuperscript{2}Guangxi Key Laboratory of Multimedia Communications and Network Technology, Nanning 530004, China.\\
		\textsuperscript{3}School of Computer Science and Engineering, Central South University, Changsha 410083, China.\\
		\textsuperscript{4}Center for AI Research \& Department of Information and Communication Technology, University of Agder, Kristiansand 4604, Norway.\\
		\textsuperscript{5}Qatar Center for Quantum Computing, College of Science and Engineering, Hamad Bin Khalifa University, Doha, Qatar.\\
		\textsuperscript{6}Faculty of Computers and Artificial Intelligence, Hurghada University, Hurghada 83523, Egypt.\\
		\textsuperscript{7}Department of Information Systems, University of Maryland, Baltimore County (UMBC), Baltimore, MD 21250, USA.\\
		*Corresponding authors.
	}
}
\begin{document}
\maketitle

\begin{abstract}
\noindent A quantum memory can reduce exponentially the number of experiments needed to learn a physical process, but real memories are noisy. Here we show that, for Pauli memory noise with eigenvalues $\lambda_X,\lambda_Y,\lambda_Z$, the advantage is governed by $\lambda_X^2+\lambda_Y^2+\lambda_Z^2$. We prove that no protocol with an entangled probe gains an exponential advantage when this sum is at most one, reduce the general case to the additivity of a Rényi-2 coherent information, and find coded memories that retain a growing advantage within 1\% of this threshold. On a superconducting processor, nine simultaneously operated Bell pairs show an advantage that grows or shrinks exponentially with system size, reversing where a parameter-free model predicts. A three-qubit repetition code turns a shrinking advantage into a growing one with all processor and readout errors included, and under dephasing its memory performs better than Bell sampling in any memory frame on a perfect processor.
\end{abstract}

\section*{Introduction}

Learning an unknown quantum process from experiments is a basic task in physics, from characterizing the noise of a quantum computer\cite{eisert2020quantum,flammia2020efficient,harper2020efficient,flammia2021pauli} to probing an unknown interaction. Randomized measurements already make many such tasks efficient\cite{huang2020predicting}, yet a striking recent insight is that the resources needed depend on whether the experimenter can keep quantum data in a quantum memory between uses of the process. For learning the Pauli eigenvalues of an $n$-qubit channel, or identifying which Pauli operator a channel encodes, any strategy that measures each output immediately needs a number of channel uses exponential in $n$, whereas a strategy that sends half of a Bell pair through the channel and measures it jointly with the half kept in memory needs only polynomially many\cite{huang2021information,chen2022exponential,aharonov2022quantum,chen2022pauli,chen2024tight,bubeck2020entanglement}; the joint measurement is a form of Bell sampling\cite{montanaro2017learning,gross2021schur}. Such separations have been observed on superconducting processors\cite{huang2022quantum,seif2026}, on a photonic platform\cite{liu2025photonic} and for bosonic channels\cite{oh2024entanglement}.

These separations assume an ideal memory. In any experiment the memory decoheres while the probe is in the channel, and a sufficiently noisy memory is no better than none: if the memory noise breaks entanglement, every protocol can be simulated by one without memory\cite{horodecki2003entanglement}. Between this limit and a perfect memory it is not known where the advantage disappears, or whether quantum error correction can move that point. Recent work has begun to address the role of noise. Cotler, Gong and Kannan\cite{cotler2026noisy} showed that noise on the coupling between the learner and the system limits the advantage and identified a critical depolarizing noise rate near $0.2$, set by the crossing of a lower and an upper bound. Seif \emph{et al.}\cite{seif2026} demonstrated entanglement-enhanced learning with noisy auxiliary qubits on a superconducting processor, with hypothesis tests on up to 64 qubits, and argued that noisy Bell pairs keep their advantage as long as a per-pair fidelity factor exceeds $1/2$. What has been missing is a converse statement, namely a memory noise level beyond which no protocol, however it encodes its memory, gains an exponential advantage, together with an experiment in which the memory noise is varied across the point where the advantage stops growing with system size.

The question parallels a central one in quantum metrology. There\cite{giovannetti2011advances,degen2017quantum}, the Heisenberg scaling of a noisy sensor can be restored by error correction\cite{kessler2014quantum,dur2014improved,arrad2014increasing} if and only if the signal Hamiltonian lies outside the span of the noise Kraus operators\cite{demkowicz2017adaptive,zhou2018achieving}; otherwise noise reduces the scaling to a constant-factor improvement\cite{escher2011general,demkowicz2012elusive,zhou2021asymptotic}. For learning with a noisy memory the boundary has a different character: it is set by a collision entropy of the memory noise, and coding moves protocols towards it.

Here we establish this boundary and test it on hardware. For memory qubits subject to a Pauli channel with eigenvalues $\boldsymbol\lambda=(\lambda_X,\lambda_Y,\lambda_Z)$, we prove that the exponential advantage disappears at $\lambda_X^2+\lambda_Y^2+\lambda_Z^2=1$ for every protocol with an entangled probe, show that it is then set by the arithmetic mean of four letter factors whereas Bell sampling achieves only their geometric mean, and reduce the general case to a single matrix inequality, which we verify numerically for all protocols with up to four memory qubits and for random stabilizer codes on up to seven. Coded memories with syndrome-aware decoding close the gap between the two means almost entirely. On the Wukong 180 superconducting processor we operate nine Bell pairs simultaneously and observe that the joint advantage grows or shrinks exponentially with the number of sites, reversing where a model without fitted parameters predicts; we place the processor on the threshold diagram; and we show that a three-qubit repetition code turns a shrinking advantage into a growing one end to end, with a memory that, under dephasing, outperforms Bell sampling in any memory frame on a perfect processor. The gain requires strongly biased noise, which we quantify.

\section*{Results}

\subsection*{Learning with a noisy quantum memory}

We consider the task of identifying an unknown $n$-qubit Pauli operator $P$ from uses of the channel $\Lambda_P(\rho)=\bigl(I+\tr(P\rho)P\bigr)/2^n$ (equivalently, learning which Pauli eigenvalue of a Pauli channel is non-zero), with $P$ drawn uniformly. In each round the learner prepares a state of the system and of an $m$-qubit memory, sends the system through $\Lambda_P$ while each memory qubit undergoes the Pauli channel $\mathcal N_{\boldsymbol\lambda}$, and measures system and memory jointly (Fig.~1a). Rounds may be adaptive; quantum data are not kept from one round to the next. The memory may encode the probe in any way, with any $m$. State preparation and measurement are taken to be ideal, which makes the converse bounds below the strongest possible.

Without memory, identifying $P$ needs $\Omega(2^n)$ channel uses\cite{chen2022exponential,chen2024tight}. With memory, the information gained per use is controlled by a single number. Let $\rho_{RA}$ be the state of the system (here acting as the reference $R$ of the channel) and the memory $A$ after the memory noise. We show (Methods) that the $\chi^2$ divergence between the outcome distributions with and without the hidden operator, averaged over $P$, is at most $(2^nK-1)/(4^n-1)$ per use, where
\begin{equation}
K=2^{-\widetilde H^{\downarrow}_2(R|A)_\rho}
\end{equation}
and $\widetilde H^\downarrow_2$ is the sandwiched Rényi-2 conditional entropy\cite{muller2013quantum,wilde2014strong,tomamichel2014relating,tomamichel2016quantum}. For $K\le1$ this is the memoryless value $1/(2^n+1)$, and the number of uses is then $\Omega(2^n)$ as without memory. An exponential advantage therefore requires that $K$ grow exponentially with the number of memory qubits.

For Bell probes the advantage is a product over sites. A uniformly drawn $P$ has, at each site $j$, one of the four letters $I$, $X$, $Y$ or $Z$, and the Bell measurement of site $j$ reveals letter $\ell$ with a squared visibility, the letter factor $G_{j,\ell}$; the identity letter needs no signal, $G_{j,I}=1$. Identifying a typical $P$ takes $\prod_j\mu_j^{-1}$ uses up to subexponential factors, with the per-site factor
\begin{equation}
\mu_j=\bigl(G_{j,I}\,G_{j,X}\,G_{j,Y}\,G_{j,Z}\bigr)^{1/4},
\end{equation}
the geometric mean of the four letter factors, so the advantage over memoryless learning, $A_n=\prod_j2\mu_j$, grows with $n$ if and only if $\mu>1/2$ (Methods). For an ideal Bell pair $G_\ell=\lambda_\ell^2$, and on the dephasing family $\boldsymbol\lambda=(a,a,1)$ used in our experiments $\mu=a$. The converse bound concerns instead the arithmetic mean of the same four numbers: for any protocol with a Bell input $K=\kappa^n$, with
\begin{equation}
\frac\kappa2=\frac{1+\lambda_X^2+\lambda_Y^2+\lambda_Z^2}{4},
\end{equation}
so its advantage is at most $\prod_j\kappa=\prod_j2\cdot\frac\kappa2$ and can grow only if the arithmetic mean exceeds $1/2$, that is, if $\lambda_X^2+\lambda_Y^2+\lambda_Z^2>1$ (Fig.~1a). The gap between the two means is what better measurements, memory frames and codes can close.

\begin{figure}[!htbp]
\centering
\includegraphics[width=\textwidth]{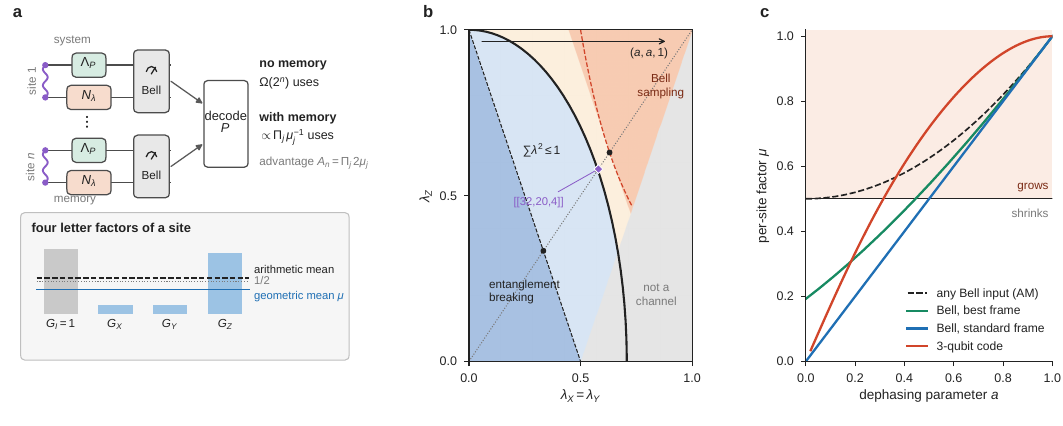}
\caption{\textbf{Where the quantum-memory advantage ends.} \textbf{a}, The learning task. At each of $n$ sites a system qubit, entangled with a memory qubit, passes through the unknown Pauli channel $\Lambda_P$ while the memory suffers the Pauli noise $N_{\boldsymbol\lambda}$; a Bell measurement of every site and joint decoding identify $P$. Without memory, $\Omega(2^n)$ uses are needed; with memory, the advantage is a product of per-site factors $2\mu_j$. Wavy lines: the Bell pair shared by system and memory qubit at each site. Each site has four letter factors, $G_I=1$ and $G_X$, $G_Y$, $G_Z$ (bars, schematic). Their arithmetic mean exceeds $1/2$ if and only if $\sum\lambda^2>1$, which bounds every protocol with a Bell input; Bell sampling achieves their geometric mean $\mu$. \textbf{b}, Pauli memory noise with $\lambda_X=\lambda_Y$. Dark blue: entanglement-breaking noise, where no protocol has an advantage. Light blue: $\sum\lambda^2\le1$, where we prove that protocols with an entangled probe have no growing advantage and conjecture the same for all protocols. Beige: advantage possible but not with Bell sampling; red: Bell sampling in the best memory frame grows ($|\lambda_X|+|\lambda_Y|+|\lambda_Z|>3\cdot2^{-2/3}$); red dashed: the same in the standard frame ($|\lambda_X\lambda_Y\lambda_Z|>1/4$). Grey: not a quantum channel. Dots on the depolarizing diagonal: $\eta=1/3$, $1/\sqrt3$ and $2^{-2/3}$; purple diamond: the $[[32,20,4]]$ code with syndrome-aware decoding, whose advantage grows down to $\eta=0.580$. Arrow: the dephasing family $(a,a,1)$ of the experiments. \textbf{c}, Per-site factor $\mu$ on the dephasing family; the advantage grows with system size where $\mu>1/2$ (shaded). Lines: the bound for any protocol with a Bell input (arithmetic mean, $(1+a^2)/2$), Bell sampling in the best memory frame ($((1+2a)/3)^{3/2}$) and in the standard frame ($a$), and the ideal three-qubit repetition code with syndrome-aware decoding.}
\end{figure}

\subsection*{A threshold at $\lambda_X^2+\lambda_Y^2+\lambda_Z^2=1$}

Our theoretical results are summarized in Fig.~1b. Write $\kappa=(1+\lambda_X^2+\lambda_Y^2+\lambda_Z^2)/2$.

\emph{Entanglement-breaking memory.} If $\mathcal N_{\boldsymbol\lambda}$ is entanglement breaking ($|\lambda_X|+|\lambda_Y|+|\lambda_Z|\le1$; for depolarizing noise with fidelity $\eta$, $\eta\le1/3$), any protocol is equivalent to a memoryless one with the same number of channel uses, because the memory channel can be written as measure-and-prepare and commutes with the unknown channel.

\emph{Entangled probes.} If the memory marginal of the input state is maximally mixed, as for every protocol that stores half of a maximally entangled state with arbitrary operations on the system side, then exactly
\begin{equation}
K=\kappa^m .
\end{equation}
Such protocols therefore have no growing advantage when $\lambda_X^2+\lambda_Y^2+\lambda_Z^2\le1$, whereas for $\lambda_X^2+\lambda_Y^2+\lambda_Z^2>1$ the bound permits one.

\emph{Reduction.} For an arbitrary protocol, $K$ depends only on the memory marginal $W$ of the input. We show that $K(W)=\tr\bigl[\bigl(R_m\,(W\otimes\mathcal N^{\otimes m}(W)^{-1/2})\bigr)^2\bigr]$, where $R_m$ is a fixed Hermitian operator on two copies of the memory that factorizes over memory qubits (Methods). The system size and the encoding have disappeared: the question is a matrix inequality for $m$-qubit density matrices. Since $K\le2^m$ for every input, the relevant statement is the value of the bound per memory qubit. For $m=1$ we solve the inequality in closed form and find $K\le\max\{1,\kappa\}$ for every input and every system size.

\emph{Stabilizer codes.} If the probe is entangled with the logical qubits of an $[[m,k]]$ stabilizer code\cite{gottesman1997stabilizer,calderbank1996good}, $K$ becomes classical: $K=2^k\cdot2^{-H_2(L|E)}$, where $H_2(L|E)$ is the conditional collision entropy of the logical error class $L$ given the syndrome $E$ of the physical Pauli error. The threshold then says that if each physical error has collision entropy of at least one bit, the logical error retains at least $k$ bits given the syndrome.

\emph{Numerics and conjecture.} Maximizing $K$ over all memory marginals, we find $\max_W K=1$ on and inside the sphere $\sum\lambda^2=1$ for all $m\le4$, for channels in several directions, and $\max_WK=\kappa^m$ outside it, with the maximum multiplicative to four or five digits for $m=2,3$ (Fig.~6a and Supplementary Fig.~S1). For random stabilizer codes with $m\le7$ (300 per $m$), $K\le1$ everywhere on and inside the sphere, even though individual syndromes exceed one (Supplementary Fig.~S2), and for CSS codes under independent bit and phase noise we reduce the statement to a single classical correlation inequality (Supplementary Section~S7). These results identify $\max_WK_m=\max\{1,\kappa\}^m$ as the general law, which we state as a conjecture:

\begin{conjecture}
For every $m$ and every $m$-qubit memory marginal $W$, $K_m(W)\le\max\{1,\kappa\}^m$. Equivalently, the Rényi-2 coherent information $-\widetilde H^\downarrow_2$ of a Pauli memory channel is additive.
\end{conjecture}

This is a new additivity property. The von Neumann coherent information, which determines the quantum capacity\cite{schumacher1996quantum,lloyd1997capacity,devetak2005private}, is superadditive for depolarizing and related channels\cite{divincenzo1998quantum,smith2007degenerate,leditzky2018dephrasure}, and multiplicativity results for output norms\cite{king2003capacity} and completely bounded norms\cite{devetak2006multiplicativity} concern the optimized ($\uparrow$) conditional entropy; the Rényi-2 quantity that governs learning with memory is the $\downarrow$ version, and it is additive in every case we have tested.

\emph{How close protocols come.} For depolarizing memory noise the threshold is $\eta=1/\sqrt3\approx0.577$. Bell sampling grows only for $\eta>2^{-2/3}\approx0.630$ ($\sum\lambda^2=1.19$), but coded memories with syndrome-aware decoding close the gap almost entirely: high-rate codes of distance 2 to 4 grow down to $\eta=0.58$--$0.60$, and the $[[32,20,4]]$ code down to $\eta=0.580$, where $\sum\lambda^2=1.010$ (Fig.~6b and Supplementary Section~S4). At $\eta=1/\sqrt3$ every code we evaluated has $\mu<1/2$, the best within 1\%, as the conjecture requires. Along the dephasing family, $\sum\lambda^2=1+2a^2>1$ for every $a>0$, so the conjectured limit lies at $a=0$, whereas Bell sampling grows for $a>1/2$ in the standard frame and for $a>(3\cdot2^{-2/3}-1)/2\approx0.445$ in the best memory frame, in which a rotation of the memory equalizes the three letter visibilities at $\sum_\ell|\lambda_\ell|/3$ (Methods); an ideal three-qubit phase-flip repetition code grows for $a>0.31$ (Fig.~1c). For $a\gtrsim0.37$ this code even exceeds the bound for any protocol with a Bell input, which it is not subject to, because its memory marginal is maximally mixed only on the code space.

\subsection*{Exponential scaling on a superconducting processor}

We implemented Bell-probe learning on the Wukong 180 processor with up to nine sites operated simultaneously (18 qubits). At each site a Bell pair is prepared; the system qubit receives a random Pauli twirl compatible with a randomly chosen letter $X$, $Y$ or $Z$, the memory qubit receives a Pauli error drawn from the dephasing channel $(a,a,1)$, and the pair is measured in the Bell basis. Letters, twirls and injected errors are drawn independently at every site in every circuit, so a single data set contains the statistics of all Pauli operators of weight $w=1,\dots,9$, from which the advantage for a uniformly drawn $P$ follows (Methods). Seven noise settings ($a=1.00$, $0.40$, $0.56$, $0.64$, $0.72$, $0.80$ and $0.88$), with 4,000 circuits of 40 shots each, were interleaved in random order; the noiseless setting serves as a running calibration. Memory noise is injected as random Pauli gates, so its eigenvalues are known exactly; the processor adds its own noise on top.

Two observations shaped the experiment. First, the calibration data supplied with the processor did not predict which qubits performed well: of twelve four-qubit chains selected by their calibrated fidelities, seven contained a faulty or strongly miscalibrated qubit. We therefore measured every candidate pair directly and selected nine pairs that are disjoint and not connected by any coupler (Supplementary Fig.~S5). Second, the platform's built-in readout correction returns rounded quasi-probabilities with negative values removed, which inflates multi-qubit correlations: with it, the joint signal of nine sites exceeded the product of single-site signals by 39\% in the noiseless setting (Supplementary Fig.~S4). The nine-pair results therefore use raw counts, for which this excess vanishes.

Figure~2a shows the joint advantage $\log_2A_w$, estimated from the product of per-site outcomes in the same shots and averaged over subsets of $w$ sites. At every noise level it is linear in $w$. Its slope changes monotonically from $-0.56$ bits per site at $a=0.40$ to $+0.49$ bits per site at $a=0.88$ and crosses zero at $a^*=0.636$. At $w=9$ the advantage is $2^{-1.22}$ at $a=0.56$ and $2^{+4.37}$ at $a=0.88$, with 95\% confidence intervals $[-1.63,-0.87]$ and $[+4.23,+4.51]$ in $\log_2$.

The reversal point is predicted without fitting. For ideal Bell pairs on this family $\mu=a$. If the processor contributes a constant factor $\mu_j(1)$ at each site, measured directly by the interleaved noiseless circuits (geometric mean $0.786$), then $\mu_j(a)=\mu_j(1)\,a$, which places the reversal at $a^*=0.5/0.786=0.636$ (Fig.~2b), in agreement with the measured value. The data follow this model letter by letter (Supplementary Fig.~S9), with a root-mean-square deviation of 3\% at eight of the nine sites, and the agreement of the reversal point holds under every variation of the analysis we tried, which gives values between $0.618$ and $0.646$, each within $0.006$ of the corresponding prediction (Fig.~2d and Supplementary Section~S12). In simulated scans that follow the model exactly, $a^*$ has a standard deviation of $0.0013$, and the measured values for different random seeds of the analysis lie within this distribution (Fig.~2e). In an independent run on a partly different set of qubits, a free power-law fit of $\nu=\mu^{4/3}$ gives an exponent of $1.34$, compared with $4/3$ (Supplementary Section~S14).

The joint statistics factorize. The ratio $R_w$ of the joint signal to the product of single-site marginals (Methods) is consistent with one for all noise levels and all $w\le9$ (Fig.~2c); in the noiseless setting $R_9=1.011$ $[0.990,1.031]$. Joint and marginal slopes agree to within $0.02$ bits per site. Splitting the run into its 187 submission batches, the site visibilities show no common drift (mean cross-site correlation $+0.01$; Supplementary Fig.~S6). Eighteen qubits operated together thus behave as nine independent sites, and the multi-site advantage is fixed by the single-site factor $\mu$.

\begin{figure}[t]
\centering
\includegraphics[width=\textwidth]{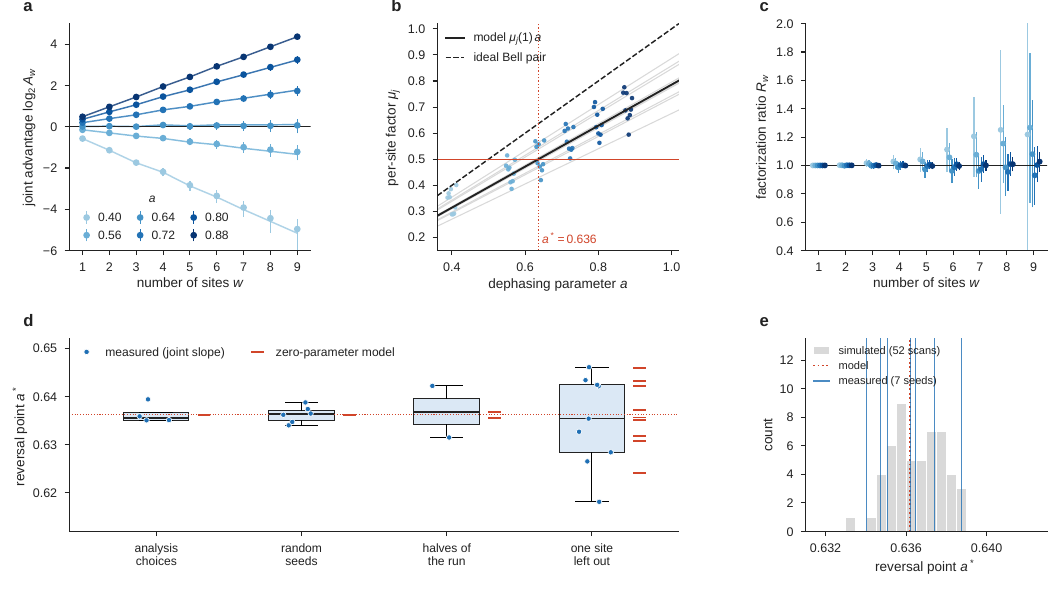}
\caption{\textbf{Exponential scaling of the advantage with system size on Wukong 180.} Nine Bell pairs (18 qubits) operated simultaneously; injected memory noise $\boldsymbol\lambda=(a,a,1)$; raw counts. \textbf{a}, Joint advantage over memoryless learning, $\log_2A_w$, for a uniformly drawn Pauli operator on $w$ sites. Points: estimate from the product of per-site outcomes in the same shots, averaged over up to 30 subsets of $w$ sites (95\% bootstrap intervals over circuits, 2,000 resamplings); lines: product of single-site marginals. \textbf{b}, Per-site factors $\mu_j$ (points, slightly displaced horizontally) and the parameter-free model $\mu_j(1)\,a$ (grey lines, one per site; black: geometric mean), with $\mu_j(1)$ from interleaved noiseless circuits. Dashed: ideal Bell pair. The dotted line marks the predicted reversal, $a^*=0.636$. \textbf{c}, Factorization ratio $R_w$ (joint signal divided by the product of marginals; points displaced horizontally for legibility; same colours as in \textbf{a}). At $a=0.40$ the joint signal for $w\ge6$ is comparable to its statistical error. \textbf{d}, Reversal point $a^*$ (zero of the joint slope) under 21 variations of the analysis: without frame alignment, without cross-fitting, with all subsets, with seven random seeds, separately for the two halves of the run in submission order, and with each site left out (points; boxes: median and quartiles), with the zero-parameter prediction for each variant (red bars) and for the full data (dotted). \textbf{e}, Sampling distribution of $a^*$ in 52 simulated scans that follow the zero-parameter model exactly (grey), the model value (dotted) and the measured values for seven random seeds of the analysis (blue).}
\end{figure}

\subsection*{The processor on the threshold diagram}

The dephasing family never reaches the threshold for an ideal memory, since $\sum\lambda^2=1+2a^2$. A real memory, however, also suffers the processor's own errors. For Bell probes the total channel of a site is measured directly: the three letter factors are its squared effective Pauli eigenvalues, so $\Sigma_j=G_{X}+G_{Y}+G_{Z}$ is the effective $\sum\lambda^2$ of site $j$, processor included (Methods). Because $\Sigma_j$ includes errors on the system qubit as well, it is a lower bound on the effective $\sum\lambda^2$ of the memory alone. On the following day we repeated the nine-pair experiment with depolarizing noise $(\eta,\eta,\eta)$, $\eta=0.55$--$0.94$, and with strong dephasing, $a=0.20$--$0.34$ (Supplementary Table~S1). The sites had not changed: the noiseless factors agreed with the previous day to within $0.05$.

Figure~3 places all 19 settings of the two days on the threshold diagram. For every site and setting the per-site factor lies, as it must, below the arithmetic-mean bound $\mu_j\le(1+\Sigma_j)/4$, and the growth rate of the joint advantage lies below $\log_2[(1+\bar\Sigma)/2]$, the largest rate available to any protocol with a Bell input (Fig.~3a,b). Five settings bring the site-averaged effective channel below $\bar\Sigma=1$ ($\bar\Sigma=0.72$--$0.99$), and in all of them the joint advantage shrinks with system size, by $0.33$--$1.46$ bits per site. Beyond $\bar\Sigma=1$ the growth rate rises smoothly and crosses zero at $\bar\Sigma=1.22$ for depolarizing noise, close to $1.19$ for ideal Bell pairs, and at $1.38$ for dephasing. In the depolarizing scan the reversal lies at $\eta^*=0.735$, against $0.739$ predicted without fitting from the noiseless factors, $\mu_j=\mu_j(1)\,\eta^{3/2}$; the per-site factors follow this prediction to within 5\% at every noise level and within 2\% for $\eta\ge0.70$ (Fig.~3c). Between $\bar\Sigma=1$ and the reversal of Bell probes, a growing advantage is permitted, but only for protocols that do better than Bell sampling.

\begin{figure}[t]
\centering
\includegraphics[width=\textwidth]{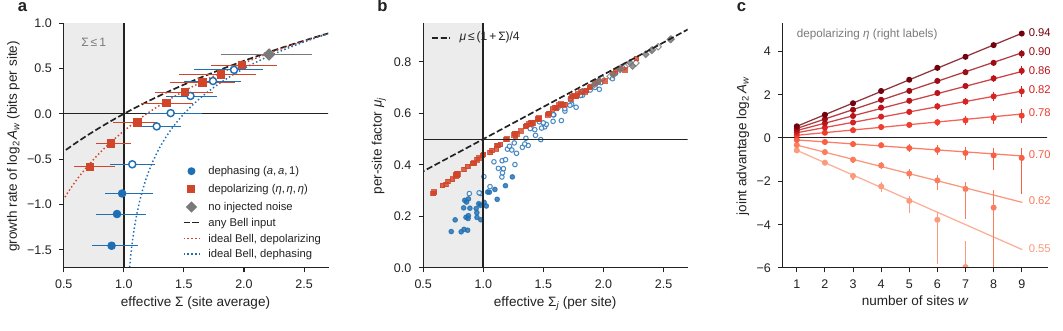}
\caption{\textbf{The processor on the threshold diagram.} Nine Bell pairs, raw counts; open symbols: dephasing $(a,a,1)$ (first day, Fig.~2); filled: depolarizing $(\eta,\eta,\eta)$ and strong dephasing (second day); grey diamonds: no injected noise. The effective $\Sigma=G_X+G_Y+G_Z$ of each site is its measured $\sum\lambda^2$ including all processor errors. \textbf{a}, Growth rate of the joint advantage (slope of $\log_2A_w$ in $w$) against the site-averaged $\Sigma$ (bars: range over sites). Grey: $\Sigma\le1$. Dashed: $\log_2[(1+\Sigma)/2]$, the largest growth rate of any protocol with a Bell input; dotted: ideal Bell pairs under depolarizing (red) and dephasing (blue) noise. \textbf{b}, Per-site factor $\mu_j$ against $\Sigma_j$ for every site and setting, with the bound $\mu\le(1+\Sigma)/4$. \textbf{c}, Joint advantage $\log_2A_w$ in the depolarizing scan, labelled by $\eta$ (95\% bootstrap intervals; intervals wider than six units omitted), with the product of single-site marginals (lines).}
\end{figure}

\subsection*{A coded memory on hardware}

A memory code can protect the probe against the dominant component of the noise. Under dephasing the $X$ and $Y$ components decay, and a three-qubit phase-flip repetition code protects them by detecting phase flips. We compared, on one four-qubit chain, an unencoded Bell pair on its first two qubits with a three-qubit code on the whole chain, interleaved in the same runs, with readout-calibration circuits interleaved as well (Methods). The cloud service does not allow mid-circuit measurement, so the syndrome is read out together with the data qubits at the end of each circuit. We report two levels: end to end, from raw counts, which includes every processor and readout error; and the memory level, after linear and unbiased readout mitigation, which isolates the memory.

End to end, the code turns a shrinking advantage into a growing one (Fig.~4a). Its per-site factor decays more slowly with the injected noise than that of the unencoded pair, so that, although it starts lower in the noiseless circuits, it crosses $\mu=1/2$ at a smaller $a$: at $0.576$ $[0.559,0.590]$ instead of $0.664$ $[0.648,0.680]$ on the second day, and at $0.569$ instead of $0.630$ and $0.562$ instead of $0.645$ in two runs on the first day. The shifts, $0.088$ $[0.066,0.111]$, $0.061$ $[0.041,0.078]$ and $0.084$ $[0.056,0.141]$, are all significant; every crossing except one (0.645, just above the largest $a=0.64$ of that run) lies within the measured range of $a$. Between the two crossings the advantage of the coded memory grows exponentially with system size while that of the unencoded memory shrinks, with all errors of the processor included.

At the memory level, the code can be compared with the best that Bell sampling can do (Fig.~4b--d). Rotating the memory qubit before storage and back before the Bell measurement can equalize the three letter visibilities; in the best frame, Bell sampling reaches $\mu=((1+2a)/3)^{3/2}$ on a perfect processor, and processor errors that act as qubit channels on the stored qubit can only lower it (Methods). In a third session we implemented this frame on the same chain. The rotation equalized the three letter factors of the pair ($0.39$, $0.42$ and $0.39$ at $a=0.50$), and the pair followed the prediction of the best-frame model without fitting to within 6\%, below the ideal best-frame curve at every $a$, as required. The coded memory, by contrast, lies above the ideal best-frame curve of a perfect processor at all twelve measured settings with $a\le0.64$ in the three sessions, significantly so at eight of them; the weighted mean excess is $0.016$ $[0.012,0.020]$ (Fig.~4d). Under dephasing, the coded memory on this processor therefore retains more information per site than Bell sampling in any memory frame would on a perfect processor. For larger $a$, where the injected noise is weak and the processor's own errors dominate, the unencoded pair is better.

\begin{figure}[t]
\centering
\includegraphics[width=\textwidth]{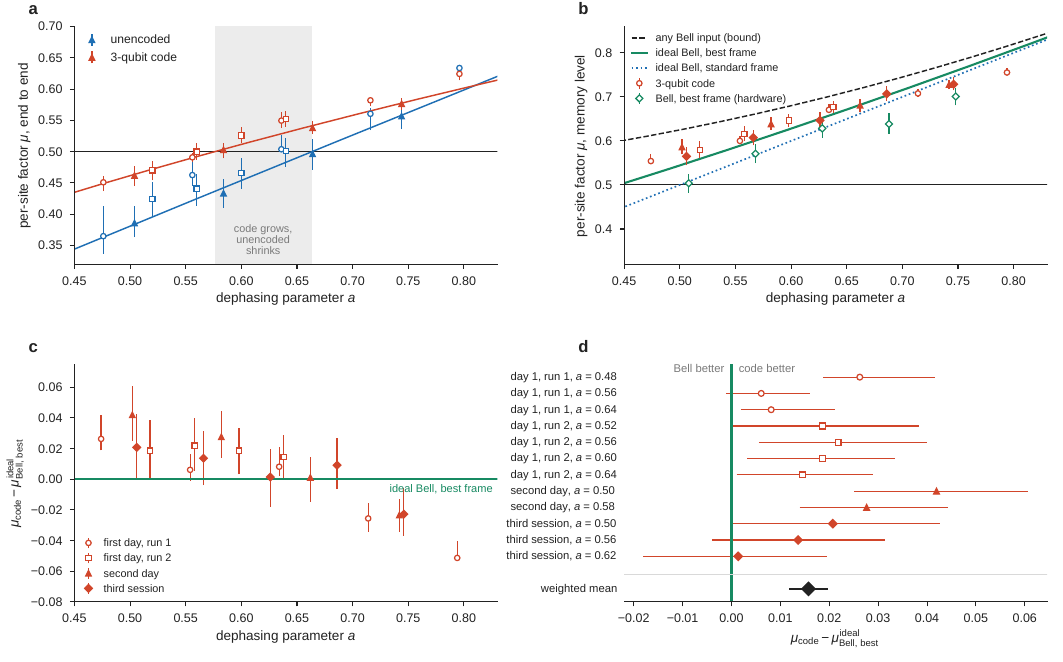}
\caption{\textbf{A coded memory on hardware.} One four-qubit chain of Wukong 180; injected memory noise $(a,a,1)$; per-site factor of an unencoded Bell pair (blue) and of a three-qubit phase-flip repetition code with syndrome-aware decoding (red). Open circles and squares: first day, runs 1 and 2; triangles: second day; diamonds: third session; bars: 95\% bootstrap intervals over circuits. \textbf{a}, End to end (raw counts, all processor and readout errors included). Lines: power-law fits $\mu=c\,a^p$ to the second day. Shaded: range in which the advantage of the coded memory grows and that of the unencoded memory shrinks (second day). \textbf{b}, Memory level (linear readout mitigation). Lines: Bell sampling on a perfect processor in the best (green) and the standard (blue dotted) memory frame, and the bound for any protocol with a Bell input (dashed). Green diamonds: the unencoded pair in the best memory frame on the same chain (third session). \textbf{c}, Per-site factor of the coded memory minus that of ideal Bell sampling in the best frame at the same $a$. \textbf{d}, The same difference for every setting with $a\le0.64$ (95\% intervals) and its inverse-variance weighted mean (black diamond, $0.016$ $[0.012,0.020]$).}
\end{figure}

\subsection*{Coding and the bias of the noise}

Theory predicts the code to keep a gain when the noise is not purely dephasing, $\boldsymbol\lambda=(t,t,\lambda_Z)$ with $\lambda_Z<1$, because the code still corrects the dominant phase flips: for an ideal processor, the coding gain $\log_2(\mu_{\rm code}/\mu_{\rm unencoded})$ at fixed $t$ decreases from $0.38$ at $\lambda_Z=1$ but stays positive down to $\lambda_Z=0.88$ (Fig.~5b). On hardware the gain depends much more sharply on the bias (Fig.~5). Averaged over the three noise levels $t=0.58$, $0.66$ and $0.74$ measured in every scan, it is $+0.162$ $[+0.119,+0.206]$ at $\lambda_Z=1$ and $+0.105$ $[+0.062,+0.149]$ at $0.98$, changes sign near $\lambda_Z\approx0.96$, and is $-0.045$ $[-0.091,+0.002]$ at $0.95$ and $-0.18$ $[-0.23,-0.13]$ at $0.90$ at the memory level; end to end the values are $+0.127$, $+0.068$, $-0.074$ and $-0.214$, with the same change of sign.

The unencoded memory follows the zero-parameter model in every noise direction, to within 5\%. For the coded memory we identify an additional mechanism: at every noisy setting of the second day its factor lies $7$--$14\%$ below the ideal value multiplied by its noiseless factor, because processor errors that occur together with injected errors produce syndromes that the decoder attributes to the injected noise. A model with this loss as a single constant describes the dependence on bias (Fig.~5b, dotted). A code therefore helps on hardware when it is matched to strongly biased noise, with a tolerance to bias that grows as the processor's own error rate falls.

\begin{figure}[t]
\centering
\includegraphics[width=\textwidth]{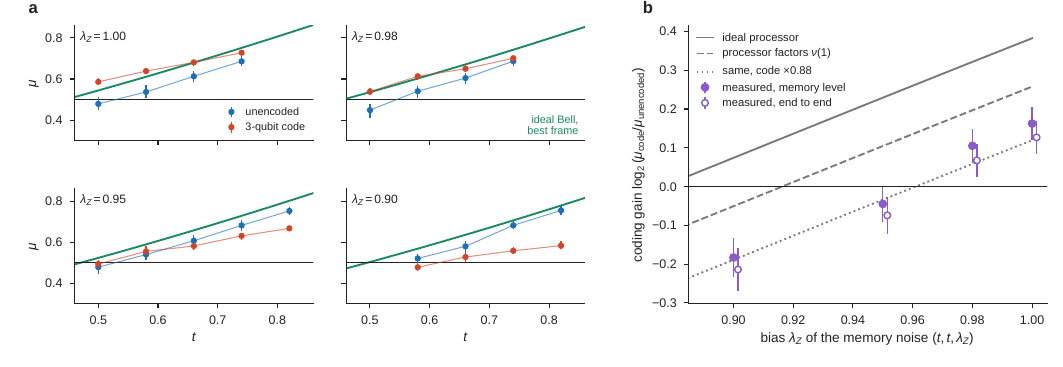}
\caption{\textbf{The coding gain depends on the bias of the noise.} Same chain, second day; injected noise $\boldsymbol\lambda=(t,t,\lambda_Z)$ on every memory qubit; 4,000 circuits per point. \textbf{a}, Per-site factor of the unencoded pair (blue) and the repetition code (red) at the memory level for $\lambda_Z=1.00$, $0.98$, $0.95$ and $0.90$ (bars: 95\% bootstrap intervals), with Bell sampling in the best memory frame on a perfect processor (green). \textbf{b}, Coding gain $\log_2(\mu_{\rm code}/\mu_{\rm unencoded})$, averaged over $t=0.58$, $0.66$ and $0.74$, at the memory level (filled) and end to end (open), with 95\% intervals. Lines: prediction for an ideal processor (solid); with the processor factors of the noiseless circuits (dashed); and with an additional, constant 12\% loss of the coded memory in noisy circuits (dotted).}
\end{figure}

\begin{figure}[t]
\centering
\includegraphics[width=0.72\textwidth]{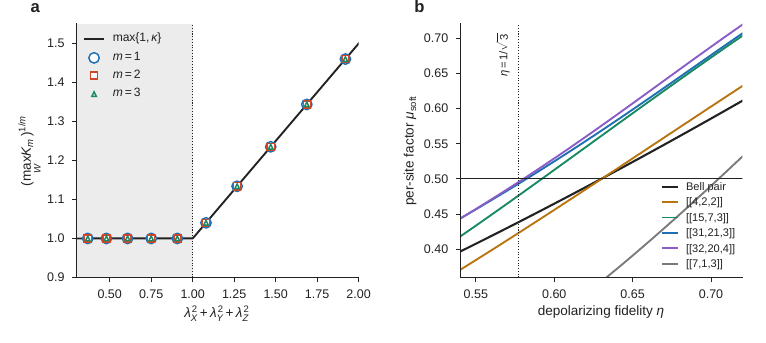}
\caption{\textbf{Evidence for the threshold beyond Bell probes.} \textbf{a}, Per-qubit maximum of $K$ over all $m$-qubit memory marginals $W$, and hence over all protocols with $m$ memory qubits, any system size and any encoding, for depolarizing noise and $m=1$ (circles), $2$ (squares) and $3$ (triangles), against $\max\{1,\kappa\}$ (line). On and inside the sphere $\sum\lambda^2=1$ (grey) the maximum is one; outside it equals $\kappa^m$, the value of Bell probes. \textbf{b}, Per-site factor $\mu_{\rm soft}$ per logical qubit of coded memories with syndrome-aware decoding under depolarizing noise, computed exactly by Walsh--Hadamard transforms over the stabilizer group, against unencoded Bell pairs ($\mu=\eta^{3/2}$, black). Dotted: the threshold $\eta=1/\sqrt3$. The $[[32,20,4]]$ code crosses $\mu=1/2$ at $\eta=0.580$; single-logical-qubit codes such as $[[7,1,3]]$ are worse than Bell pairs because the factor is counted per logical qubit.}
\end{figure}

\section*{Discussion}

The advantage of a quantum memory for learning Pauli channels survives memory noise up to $\lambda_X^2+\lambda_Y^2+\lambda_Z^2=1$, a boundary that we have proved for entangled probes and conjecture for all protocols. The quantity that decides it is a collision entropy: the threshold is where the Pauli error suffered by each memory qubit reaches a collision entropy of one bit ($2\sum_ap_a^2=\kappa=1$). In terms of the four letter factors of a site, the threshold is set by their arithmetic mean and Bell sampling by their geometric mean, and the protocols we studied interpolate between the two: rotating the memory frame equalizes the letter factors, and codes with syndrome-aware decoding come within 1\% of the threshold for depolarizing noise. This contrasts with metrology, where a span condition decides whether coding can restore the scaling\cite{zhou2018achieving}; here coding moves protocols towards a boundary fixed by the memory noise alone.

On the experimental side, the effective memory channel measured by the letter factors places a processor directly on the threshold diagram. On Wukong 180, Bell probes reverse at an effective $\sum\lambda^2\approx1.2$ for depolarizing noise and $1.4$ for dephasing, and the processor-specific factor $\mu_j(1)$ predicts the reversal for both noise directions without fitting. The factorization of the joint statistics shows that, for 18 qubits operated together, the advantage of simultaneously operated sites is fixed by single-site properties. A three-qubit code already turns a shrinking advantage into a growing one end to end, and at the memory level it performs better under dephasing than any frame of Bell sampling on a perfect processor; its gain depends on the bias of the noise, because processor errors enter the syndromes the code relies on. Coded memories with lower processor error rates and mid-circuit syndrome extraction should move the reversal further towards the threshold, as the ideal three-qubit code ($\sum\lambda^2=1.20$) shows. The experiments also suggest two practical rules for such work: select qubits by direct measurement rather than by calibration data, and use raw counts or unbiased readout mitigation rather than corrections that round and truncate quasi-probabilities, which create spurious multi-qubit correlations.

The remaining step towards a complete theory is to prove the additivity of the Rényi-2 coherent information of Pauli channels. For stabilizer codes it reduces to a classical statement about collision entropies of logical errors. For CSS codes under independent bit and phase noise, which include the dephasing family of our experiments, we reduce it further to a single classical correlation inequality of Griffiths--Ginibre type for product measures, which holds in all numerical tests and in computer-generated certificates (Supplementary Section~S7). Our converse covers protocols that do not keep quantum data from one use to the next, which suffices for Bell probes, where one use per round already achieves the ideal scaling; extending it to protocols that store quantum data across uses, to non-Pauli memory noise such as amplitude damping, and to learning tasks beyond Pauli channels are natural next steps.

\section*{Methods}

\subsection*{Learning task and protocols with a noisy memory}
Let $d=2^n$ and let $\mathcal P_n$ be the $d^2-1$ non-identity $n$-qubit Pauli strings. For a hidden $P\in\mathcal P_n$, drawn uniformly, and sign $s=\pm1$ the channel is
\begin{equation}
\Lambda_{(P,s)}(\rho)=\frac{I+s\,\tr(P\rho)\,P}{d},
\end{equation}
a uniform mixture of the $d^2/2$ Pauli operators that commute ($s=+1$) or anticommute ($s=-1$) with $P$; it is realized exactly by a random Pauli twirl\cite{wallman2016noise}. The task is to identify $P$; the null hypothesis is the completely depolarizing channel $\mathcal D(\rho)=I/d$. In each round the learner prepares a pure state $\psi$ of the system $S$ and an $m$-qubit memory $A$, applies $\Lambda_{(P,s)}$ to $S$ while every memory qubit undergoes the Pauli channel
\begin{equation}
\mathcal N_{\boldsymbol\lambda}(\rho)=\sum_{a\in\{I,X,Y,Z\}}p_a\,\sigma_a\rho\,\sigma_a,\qquad
\begin{pmatrix}p_I\\p_X\\p_Y\\p_Z\end{pmatrix}=\frac14\begin{pmatrix}1+\lambda_X+\lambda_Y+\lambda_Z\\1+\lambda_X-\lambda_Y-\lambda_Z\\1-\lambda_X+\lambda_Y-\lambda_Z\\1-\lambda_X-\lambda_Y+\lambda_Z\end{pmatrix},
\end{equation}
and measures $SA$ jointly. Rounds may be adaptive; quantum data are not kept from one round to the next (Supplementary Remark~S3.3). Writing $\rho'=(\mathrm{id}_S\otimes\mathcal N_{\boldsymbol\lambda}^{\otimes m})(\psi)$, the output of a round is
\begin{equation}
\omega_{(P,s)}=\omega_0+s\,D_P,\qquad \omega_0=\frac{I}{d}\otimes\rho'_A,\qquad D_P=\frac1d\,P\otimes\tau'_P,\qquad \tau'_P=\tr_S\bigl[(P\otimes I)\rho'\bigr].
\end{equation}

\subsection*{Detection bound}
For a measurement $\{M_x\}$ the per-use detection quantity is the classical $\chi^2$ divergence between the outcome distributions of $\omega_0+D_P$ and $\omega_0$, averaged over $P$,
\begin{equation}
G=\mathbb E_P\sum_x\frac{\bigl(\tr M_xD_P\bigr)^2}{\tr M_x\omega_0}\ \le\ \mathbb E_P\,\tr\bigl[D_P\,\omega_0^{-1/2}D_P\,\omega_0^{-1/2}\bigr]=\mathbb E_P\,\tr\bigl[\tau'_PY\tau'_PY\bigr],\qquad Y=\rho_A'^{-1/2}.
\end{equation}
The inequality is the monotonicity of the quantum $\chi^2$ divergence $\chi^2_{1/2}(\rho,\sigma)=\tr[(\rho-\sigma)\sigma^{-1/2}(\rho-\sigma)\sigma^{-1/2}]$ under quantum channels\cite{temme2010chi,petz1996monotone}. With $\sum_{P}P\otimes P=d\,\mathrm{SWAP}$ (sum over all $d^2$ Pauli strings) and $\tr(\tau'_IY\tau'_IY)=1$,
\begin{equation}
\sum_{P}\tr\bigl[\tau'_PY\tau'_PY\bigr]=d\,K(\rho'),\qquad G\le\frac{d\,K(\rho')-1}{d^2-1},
\end{equation}
with $K$ as in equation~(1). Without memory $K=\tr\psi_S^2=1$ and $G\le1/(d+1)$. On every leaf of an adaptive protocol tree the likelihood ratio is $\prod_t(1+s\,y_t)$ with $y_t=\tr(M_{x_t}D_P)/\tr(M_{x_t}\omega_0)$; averaging over $s$ and $P$ under the null distribution gives $\mathrm{TV}(\mathbb E_{P,s}p_{(P,s)},p_0)\le M\,G_{\max}$ after $M$ uses. Appending a two-use parity test, which succeeds with probability one if $\hat P=P$ and $1/2$ otherwise, yields (Supplementary Theorem~S3.1)
\begin{equation}
\Pr[\hat P=P]\le2(M+2)\,\frac{d\,K_{\max}-1}{d^2-1},\qquad M_{0.9}\ge0.45\,\frac{d^2-1}{d\,K_{\max}-1}-2,
\end{equation}
where $K_{\max}$ is the largest $K$ over the inputs a protocol uses and $\Pr[\hat P=P]$ is averaged over the uniform $P$. For $K_{\max}\le1$ this is the memoryless bound $0.45(d+1)-2$.

\subsection*{Entanglement-breaking memories}
If $\mathcal N_{\boldsymbol\lambda}$ is entanglement breaking, so is $\mathcal N_{\boldsymbol\lambda}^{\otimes m}$, which can be written as $\mathcal M(X)=\sum_k\tr(F_kX)\tau_k$. Because $\Lambda_{(P,s)}$ acts on $S$ only, the round output is $\sum_k\Lambda_{(P,s)}(\sigma_k)\otimes\tau_k$ with $\sigma_k=\tr_A[(I\otimes F_k)\psi]$, which a memoryless protocol reproduces by sampling $k$ with probability $\tr\sigma_k$, sending $\sigma_k/\tr\sigma_k$ and measuring with $M^{(k)}_x=\tr_A[(I\otimes\tau_k)M_x]$.

\subsection*{Entangled probes}
Expand the noisy state in memory Pauli strings $\tau$ as $\rho'=2^{-m}\sum_\tau\lambda_\tau M_\tau\otimes\tau$ with $M_\tau=\tr_A[(I\otimes\tau)\psi]$ and $\lambda_\tau=\prod_j\lambda_{\tau_j}$ ($\lambda_I=1$). If the memory marginal is maximally mixed, then $\rho'_A=I/2^m$, all Schmidt coefficients equal $2^{-m}$ and $\|M_\tau\|_2^2=2^{-m}$ for every $\tau$, so
\begin{equation}
K=2^m\tr\rho'^2=2^{-m}\sum_\tau\lambda_\tau^2=\Bigl(\frac{1+\lambda_X^2+\lambda_Y^2+\lambda_Z^2}{2}\Bigr)^{m}=\kappa^m .
\end{equation}

\subsection*{Reduction to the memory marginal}
Write $\psi=(X\otimes I)|\Gamma\rangle\langle\Gamma|(X^\dagger\otimes I)$ with $|\Gamma\rangle=\sum_i|ii\rangle$ and $X^\dagger X=W^{\mathsf T}$, where $W=\psi_A$. Then $\rho'_A=\mathcal N^{\otimes m}(W)$ and $K$ is the trace of the superoperator $\mathcal N^{\otimes m}\circ\Phi_W\circ\mathcal N^{\otimes m}\circ\Phi_Z$, with $\Phi_V(\cdot)=V\cdot V$ and $Z=\mathcal N^{\otimes m}(W)^{-1/2}$. Expanding $\mathcal N^{\otimes m}=\sum_\mu p_\mu\sigma_\mu\cdot\sigma_\mu$ ($p_\mu=\prod_jp_{\mu_j}$) and using $\mathrm{Tr}(X\mapsto LXR)=\tr L\,\tr R$ gives, for any system size and any encoding,
\begin{equation}\label{eq:red}\begin{gathered}
K(W)=\sum_{\mu,\nu}p_\mu p_\nu\,\bigl|\tr\bigl(\sigma_\mu W\sigma_\nu Z\bigr)\bigr|^2=\tr\Bigl[\bigl(R_m\,(W\otimes Z)\bigr)^2\Bigr],\\ R_m=\Bigl(\sum_\mu p_\mu\,\sigma_\mu\otimes\sigma_\mu\Bigr)\mathrm{SWAP}=R_1^{\otimes m}.
\end{gathered}
\end{equation}
Both forms agree with the definition to $10^{-15}$ on random inputs. $R_1$ is Hermitian but not positive (its smallest eigenvalue is about $-0.26$ to $-0.37$ on the threshold sphere, depending on the direction), which rules out standard trace inequalities for positive operators. The conjecture is the statement $\max_WK(W)=\max\{1,\kappa\}^m$.

\subsection*{One memory qubit}
For $m=1$ the system can be taken to be one qubit and $W=\frac12(I+r\,\hat n\cdot\boldsymbol\sigma)$. With $L=\sum_k\lambda_k^2n_k^2$, $q^2=r^2L$, $D=1-q^2$, $\Lambda_2=\sum_k\lambda_k^2$ and $M=\sum_k\lambda_k^4n_k^2/L\in[0,1]$, a direct calculation gives
\begin{equation}
2K=\frac{1+r^2-2q^2}{D}+(1-r^2)\Bigl[\frac{\Lambda_2}{\sqrt D}+M\Bigl(\frac1D-\frac1{\sqrt D}\Bigr)\Bigr],
\end{equation}
in which every coefficient is non-negative because $D\le1$. If $\Lambda_2\le1$, then $M\le1$ and
\begin{equation}
2K\le\frac{1+r^2-2r^2L}{1-r^2L}+\frac{1-r^2}{1-r^2L}=2 .
\end{equation}
If $\Lambda_2>1$, $M\le1$ gives $2K\le2+(\Lambda_2-1)(1-r^2)/\sqrt{1-r^2L}\le1+\Lambda_2$, because $(1-r^2)^2\le1-r^2L$ is equivalent to $r^2(r^2+L-2)\le0$. Hence $K\le\max\{1,\kappa\}$, with equality for the Bell input when $\Lambda_2>1$.

\subsection*{Stabilizer-code memories}
For a probe maximally entangled with the logical qubits of an $[[m,k]]$ code with projector $\Pi_C$, $W=\Pi_C/2^k$ and $\mathcal N^{\otimes m}(W)$ is diagonal in the syndrome basis, with eigenvalue $q_e/2^k$ on the sector $\Pi_e$ of syndrome $e$. The trace $\tr(\sigma_\mu\Pi_C\sigma_\nu\Pi_e)$ vanishes unless $\mu$ and $\nu$ have syndrome $e$ and belong to the same logical class $L$, in which case its modulus is $2^k$. Hence
\begin{equation}\label{eq:stab}
K=2^k\sum_e\sum_L\frac{P(e,L)^2}{q_e}=2^k\,2^{-H_2(L|E)},\qquad H_2(L|E)=-\log_2\sum_eq_e\sum_LP(L|e)^2,
\end{equation}
where $P(e,L)$ is the joint distribution of syndrome and logical class of the physical Pauli error. Since $2\sum_ap_a^2=\kappa$, the condition $\kappa\le1$ states that each physical error has collision entropy at least one bit, and the conjecture for stabilizer codes states that the logical error then retains at least $k$ bits given the syndrome.

\subsection*{Achievability and the per-site factor}
At each site a memory (one qubit, or one logical qubit of a code) is maximally entangled with system qubit $j$. Decoding yields syndromes $\mathbf s$ and a logical Pauli outcome $R'=RL$, where $R$ is the channel's Pauli and $L$ the logical memory error. Under $(P,s)$,
\begin{equation}
p_{(P,s)}(\mathbf s,R')=\frac{q(\mathbf s)}{4^n}\Bigl(1+s\,\chi_P(R')\,\mu_P(\mathbf s)\Bigr),\qquad \mu_P(\mathbf s)=\prod_j\lambda_{j,P_j,s_j},
\end{equation}
where $\chi_P(R')=\pm1$ records whether $R'$ commutes with $P$ and $\lambda_{j,\ell,s}=\mathbb E[\chi_\ell(L_j)\mid s_j=s]$, with $\lambda_{j,I,s}=1$. The Bhattacharyya coefficient\cite{kailath1967divergence} between $(P,s)$ and $(Q,s')$, $Q\ne P$, is at most $1-g_P/8$ with
\begin{equation}
g_P=\prod_jG_{j,P_j},\qquad G_{j,\ell}=\sum_sq_{j,s}\,\lambda_{j,\ell,s}^2,
\end{equation}
so maximum-likelihood decoding identifies $P$ with probability $1-\delta$ after
\begin{equation}
M\ \ge\ \frac{8}{g_P}\Bigl(\ln(2\cdot4^n)+\ln\frac1\delta\Bigr)
\end{equation}
non-adaptive uses (Supplementary Proposition~S4.2). For $P$ drawn uniformly, the letters $P_j$ are independent and uniform on $\{I,X,Y,Z\}$ (up to the excluded identity string), so $\log_2g_P$ is a sum of independent terms with mean $\sum_j\log_2\mu_j$,
\begin{equation}\label{eq:mu}
\mu_j=\bigl(G_{j,I}\,G_{j,X}\,G_{j,Y}\,G_{j,Z}\bigr)^{1/4}=\nu_j^{3/4},\qquad \nu_j=\bigl(G_{j,X}\,G_{j,Y}\,G_{j,Z}\bigr)^{1/3},
\end{equation}
and fluctuations of order $\sqrt n$. Identifying $P$ with probability $0.9$ on average therefore takes $\prod_j\mu_j^{-1}$ uses up to a factor $2^{O(\sqrt n)}$, and comparison with the memoryless bound gives $\log_2A_n=\sum_j\log_2(2\mu_j)+O(\sqrt n)$: the advantage grows with $n$ if and only if $\mu>1/2$. For unencoded pairs $G_\ell=\lambda_\ell^2$, $\mu=|\lambda_X\lambda_Y\lambda_Z|^{1/2}$, which equals $a$ on the dephasing family and $\eta^{3/2}$ for depolarizing noise. By the inequality of arithmetic and geometric means, $\mu\le(1+\lambda_X^2+\lambda_Y^2+\lambda_Z^2)/4=\kappa/2$, the per-site bound for any protocol with a Bell input. For codes under depolarizing noise, $q_s$ and $A_s(\sigma)=q_s\lambda_{\sigma,s}$ follow exactly from Walsh--Hadamard transforms over the stabilizer group $S$ with $r$ generators,
\begin{equation}\label{eq:wht}
q_s=2^{-r}\sum_{g\in S}(-1)^{s\cdot g}\,\eta^{\mathrm{wt}(g)},\qquad A_s(\sigma)=2^{-r}\sum_{g\in S}(-1)^{s\cdot g}\,\eta^{\mathrm{wt}(g\bar\sigma)},
\end{equation}
and $\mu_{\rm soft}$ is the geometric mean of $G_\sigma^{1/k}$ over logical strings $\sigma\in\{I,X,Y,Z\}^k$.

\emph{Memory frame.} If the memory is rotated by one unitary before storage and by another before the Bell measurement, the letter visibilities are the diagonal entries of $M=R\,T_2DT_1\,O$, where $D=\mathrm{diag}(\boldsymbol\lambda)$, $O,R\in SO(3)$ are the induced rotations and $T_{1,2}$ the $3\times3$ blocks of the processor operations before and after storage, modelled as qubit channels on the stored qubit (no leakage); the $3\times3$ block of any qubit channel is a contraction ($T_{1,2}=I$ on a perfect processor). For a diagonal sign matrix $S$, $\sum_\ell|M_{\ell\ell}|=\tr(SM)$ is at most the sum of the singular values of $M$, which is at most that of $D$, $|\lambda_X|+|\lambda_Y|+|\lambda_Z|$. With the inequality of arithmetic and geometric means, Bell sampling in any memory frame therefore has $\mu\le((|\lambda_X|+|\lambda_Y|+|\lambda_Z|)/3)^{3/2}$, with equality on a perfect processor (Supplementary Proposition~S4.1): a $\pi$ rotation of the memory, or the sign of the visibilities, which does not enter $G_\ell=v_\ell^2$, makes all eigenvalues non-negative, and then $O$ mapping the $Z$ axis to $(1,1,1)/\sqrt3$ and $R=O^{\mathrm T}$ equalize the visibilities when $|\lambda_X|=|\lambda_Y|$. Bell sampling therefore has a growing advantage in some memory frame if and only if $|\lambda_X|+|\lambda_Y|+|\lambda_Z|>3\cdot2^{-2/3}$; for the non-negative eigenvalues of the noise families used here the absolute values can be dropped. The third session implemented this frame with $U=R_Y(-\theta)R_Z(-\pi/4)$, $\cos\theta=1/\sqrt3$, applied to the memory qubit after the Bell pair is prepared and inverted before the Bell measurement.

\subsection*{Numerics}
For $m\le4$ we maximized equation~\eqref{eq:red} over $W=XX^\dagger/\tr XX^\dagger$ with L-BFGS after sampling 200 random $W$ per channel, for channels on the sphere $\sum\lambda^2=1$ in three directions, inside it, and outside it as a positive control (where the optimizer reaches $\kappa^m$); Fig.~6a and Supplementary Fig.~S1 show $m\le3$ along two families of channels. Random stabilizer codes were generated from random Clifford tableaux with random $k$ and evaluated exactly with equation~\eqref{eq:stab} (Supplementary Fig.~S2). Code thresholds use equation~\eqref{eq:wht} with soft decoding, averaging over all $4^k$ logical strings for $k\le4$ and over 1,000 random strings otherwise; the $[[32,20,4]]$ code is the CSS code whose $X$ and $Z$ stabilizers are both given by the generator matrix of the first-order Reed--Muller code $\mathrm{RM}(1,5)$ (Fig.~6b and Supplementary Section~S4). The ideal values of the hardware protocols were computed by enumerating all Pauli errors of the memory qubits through the encoder and decoder of the experiment.

\subsection*{Experimental details}
\emph{Device.} Experiments ran on the Wukong 180 superconducting processor (Origin Quantum), whose native gates are $R_\phi(\theta)=\exp[-i\theta(\cos\phi\,X+\sin\phi\,Y)/2]$ and CZ, accessed through its cloud service between 16 and 20 November 2026. Calibration data supplied with the processor for the qubits used are listed in Supplementary Table~S15, together with the noiseless per-site factor we measured; the supplied data were not updated during the experiments and did not predict which qubits performed well (Supplementary Section~S9).

\emph{Circuits.} At a Bell-pair site with system qubit $s$ and memory qubit $m_1$ the circuit is
\begin{equation}
U_{\rm enc}=(I\otimes H)\,\mathrm{CZ}_{s,m_1}\,(H\otimes H),\qquad
\text{circuit}=U_{\rm enc}^\dagger\;\bigl(R_s\otimes E_{m_1}\bigr)\;U_{\rm enc},
\end{equation}
followed by measurement of both qubits in the computational basis, which realizes a Bell measurement. $R_s$ is a Pauli drawn uniformly from the commutant $\{I,\ell\}$ of the site's letter $\ell\in\{X,Y,Z\}$ (the hidden channel), and $E_{m_1}$ is a Pauli drawn from the distribution of the injected memory channel $\mathcal N_{\boldsymbol\lambda}$; we used the families $(a,a,1)$ (dephasing), $(\eta,\eta,\eta)$ (depolarizing) and $(t,t,\lambda_Z)$ with $\lambda_Z=0.98$, $0.95$, $0.90$ (biased). The two outcome bits $(z,x)$ give the parities $\chi_X=(-1)^z$, $\chi_Z=(-1)^x$ and $\chi_Y=\chi_X\chi_Z$. For the repetition code the encoder entangles $s$ with the code on $(m_1,m_2,m_3)$ by a chain of three CZ gates conjugated by Hadamards, $E$ acts on all three code qubits, and after the inverse encoder $m_2,m_3$ hold the two-bit syndrome. Letters, twirls and injected errors are drawn independently for every site in every circuit, so one data set contains the statistics of all Pauli strings of weight $w=1,\dots,N$. Pauli gates are implemented as $X$ and $R_Z(\pi)$ ($Y\simeq R_Z(\pi)X$ up to phase).

\emph{Compilation.} The cloud service's circuit optimization was disabled, because it cancels gates across CZ gates and miscompiled circuits with injected Pauli errors (Supplementary Fig.~S3 and Section~S9). Every run of single-qubit gates on a qubit was compiled into one $R_\phi$ pulse, with the remaining $Z$ rotation carried as a virtual frame through CZ gates and dropped before measurement\cite{mckay2017efficient}; each compiled circuit was checked against exact simulation. Qubits were pinned to the selected sites.

\emph{Runs.} The experiments are summarized in Supplementary Table~S1. All noise settings of a run (and, in the code experiments, both memories and the readout-calibration circuits) were shuffled together and submitted in batches of 150 (first day) or 200 (later sessions) circuits, so that they share the processor's drift. Server-side readout correction was disabled. The nine-pair results use raw counts; the coding results are reported both from raw counts (end to end) and after linear readout mitigation (memory level). A test of mid-circuit measurement was rejected by the cloud service, which does not allow a qubit to be measured twice in one circuit (Supplementary Section~S9).

\subsection*{Data analysis}
\emph{Letter factors.} For site $j$, letter $\ell$ and syndrome class $s$, let $S_{c}$ be the sum of the (frame-aligned) parities over the $n_c$ shots of circuit $c$ in that class. The squared amplitude is estimated without bias from pairs of shots in different circuits,
\begin{equation}\label{eq:G}
\widehat{\lambda^2_{\ell s}}=\frac{\bigl(\sum_cS_c\bigr)^2-\sum_cS_c^2}{\bigl(\sum_cn_c\bigr)^2-\sum_cn_c^2},\qquad
\widehat G_{j,\ell}=\sum_s\frac{\sum_cn_c}{N_\ell}\,\widehat{\lambda^2_{\ell s}},
\end{equation}
where $N_\ell$ is the number of shots with letter $\ell$, and $\mu_j$ follows from equation~\eqref{eq:mu}.

\emph{Frame alignment.} The interleaved noiseless circuits give each site's $3\times3$ letter-to-parity matrix $T_{\ell k}=\langle\chi_k\rangle_\ell$. For letter $\ell$ we evaluate the parity along the unit vector $u_\ell=T_{\ell\cdot}/\|T_{\ell\cdot}\|$, keeping off-diagonal elements only when they exceed three standard errors; the magnitude is never rescaled. Switching the alignment off changes the reversal point of the nine-pair experiment by $0.004$ (Supplementary Table~S8). The quality used for site selection is $q=\prod_\ell\|T_{\ell\cdot}\|^{2/3}$.

\emph{Joint estimator and factorization.} For a subset $S$ of sites and circuit $c$ with letters $\ell_j$,
\begin{equation}
Y_c=\Bigl\langle\prod_{j\in S}\chi_j\,w_j(\ell_j,s_j)\Bigr\rangle_{\rm shots},\qquad
M_c=\prod_{j\in S}\sum_sq_{j,\ell_j,s}\,\lambda_{j,\ell_j,s}\,w_j(\ell_j,s),\qquad
R_S=\frac{\sum_cY_c}{\sum_cM_c},
\end{equation}
with weights $w_j$ equal to the syndrome-conditioned amplitudes estimated on the other half of the circuits (cross-fitting) and $q,\lambda$ in $M_c$ from the half being evaluated. $R_S=1$ for independent sites. For a uniformly drawn $P$, each site of $S$ lies in its support with probability $3/4$, and the joint advantage is
\begin{equation}
\log_2A_S=\sum_{j\in S}\log_2(2\mu_j)+\mathbb E_T\log_2R_T,
\end{equation}
where $T\subseteq S$ is the random support; $\mathbb E_T\log_2R_T$ is evaluated with binomial weights from the mean of $\log_2R$ over subsets of each size. $\log_2A_w$ and $R_w$ are averages over up to 30 random subsets of each size $w$. Intervals are 95\% percentile bootstrap intervals over circuits (2,000 resamplings); the single-site factors are re-estimated in each resample, so the intervals of $\log_2A_w$ include their uncertainty. The estimators and intervals were validated on simulated data, where the 95\% intervals of $\log_2A_9$ covered the truth in 95--99\% of 150 data sets, and the results were checked under variations of every analysis choice (Supplementary Sections~S10--S12).

\emph{Readout mitigation.} From the calibration circuits we obtain, for each qubit $i$, the confusion matrix $C_i=\begin{pmatrix}1-e^{(i)}_0&e^{(i)}_1\\e^{(i)}_0&1-e^{(i)}_1\end{pmatrix}$ with $e^{(i)}_b$ the probability of reading $1-b$ after preparing $b$. The mitigated class sums of a site are obtained by applying $\bigotimes_{i\in\text{site}}C_i^{-1}$ to each circuit's site-marginal counts,
\begin{equation}\label{eq:mit}
\tilde h_c(y)=\sum_x\Bigl[\bigotimes_iC_i^{-1}\Bigr]_{yx}h_c(x),
\end{equation}
without clipping, so that the sums entering equation~\eqref{eq:G} remain unbiased\cite{bravyi2021mitigating,nation2021scalable}. Raw readout errors $P(1|0)$ and $P(0|1)$ of the chain were $0.7$--$8.8\%$ and $1.6$--$4.4\%$ over the five calibrations of the first two days (Supplementary Fig.~S3b and Section~S16). In the third session the readout error of the system qubit had grown to $P(1|0)=0.47$, so that session is analysed at the memory level only; the mitigation is validated there by the noiseless factor of the unencoded pair, $\nu=0.912$, against $0.907$ on the previous day. The platform's own readout correction instead returns rounded quasi-probabilities with negative entries removed, which inflated the nine-site factorization ratio to $R_9=1.39$ in the noiseless setting (Supplementary Fig.~S4); for single-site quantities it agrees with equation~\eqref{eq:mit}.

\emph{Effective memory channel.} For a Bell probe with an ideal system side, $G_{j,\ell}$ is the square of the effective Pauli eigenvalue $\lambda_\ell$ of the whole memory channel of site $j$, including preparation, gate, idling and readout errors after Pauli twirling. We therefore use $\Sigma_j=G_{j,X}+G_{j,Y}+G_{j,Z}$ as the effective $\sum\lambda^2$ of the site (Fig.~3). $\Sigma_j$ also contains errors on the system qubit, which the theory does not count as memory noise; for uncorrelated system-side and memory-side Pauli errors every eigenvalue of the combined channel is the product of a system-side and a memory-side eigenvalue, so $\Sigma_j$ is then a lower bound on the effective $\sum\lambda^2$ of the memory alone. For Bell probes $\mu_j\le(1+\Sigma_j)/4$ by the inequality of arithmetic and geometric means.

\emph{Fits.} The end-to-end crossings in Fig.~4a are from weighted least-squares fits of $\log\mu=\log c+p\log a$ within the measured range of $a$, with errors inflated by the reduced $\chi^2$ when it exceeds one and intervals from a parametric bootstrap (5,000 draws). The comparisons of Fig.~4b--d and Fig.~5 use the measured points directly; the weighted mean in Fig.~4d uses weights inversely proportional to the squared half-widths of the 95\% intervals.

\section*{Code availability}

\end{document}